\documentclass[runningheads]{llncs}
\usepackage{amsmath,amsfonts,amssymb}
\usepackage{graphicx}
\usepackage[colorlinks=true, allcolors=blue]{hyperref}
\usepackage{ulem}
\usepackage{multirow}
\usepackage{algorithm}
\usepackage{algorithmicx}
\usepackage{algpseudocode}
\usepackage{xcolor}
\begin{document}
\title{Frequency Disentangled Learning for Segmentation of Midbrain Structures from Quantitative Susceptibility Mapping Data}

\author{Guanghui Fu\inst{1} \and
Gabriel Jimenez\inst{1} \and
Sophie Loizillon\inst{1} \and
Lydia Chougar\inst{1,2,3}  \and
Didier Dormont \inst{1,3}  \and
Romain Valabregue\inst{2,4}  \and
Ninon Burgos \inst{1} \and
Stéphane Lehéricy \inst{2,3}  \and
Daniel Racoceanu \inst{1} \and
Olivier Colliot*\inst{1} \and
the ICEBERG Study Group \inst{2}
}

\authorrunning{Guanghui et al.}

\institute{Sorbonne Université, Institut du Cerveau - Paris Brain Institute - ICM, CNRS, Inria, Inserm, AP-HP, Hôpital de la Pitié Salpêtrière, Paris, France \and
ICM, Centre de NeuroImagerie de Recherche-CENIR, Paris, France \and
AP-HP, Hôpital Pitié Salpêtrière, DMU DIAMENT, Dep. of Neuroradiology, Paris, France \and
Sorbonne Université, Institut du Cerveau - Paris Brain Institute - ICM, CNRS, Inserm, AP-HP, Hôpital de la Pitié Salpêtrière, Paris, France
\email{firstname.lastname@icm-institute.org}}

\maketitle

\begin{abstract}
One often lacks sufficient annotated samples for training deep segmentation models.  This is in particular the case for less common imaging modalities such as Quantitative Susceptibility Mapping (QSM).  It has been shown that deep models tend to fit the target function from low to high frequencies.  One may hypothesize that such property can be leveraged for better training of deep learning models.
In this paper, we exploit this property to propose a new training method based on frequency-domain disentanglement. It consists of two main steps:
i) disentangling the image into high- and low-frequency parts and feature learning; ii) frequency-domain fusion to complete the task. The approach can be used with any backbone segmentation network.
We apply the approach to the segmentation of the red and dentate nuclei from QSM data which is particularly relevant for the study of parkinsonian syndromes. We demonstrate that the proposed method provides considerable performance  improvements for these tasks.  We further applied it to three public datasets from the Medical Segmentation Decathlon (MSD) challenge.  For two MSD tasks,  it provided smaller but still substantial improvements (up to 7 points of Dice),  especially under small training set situations. 
The source code of the project is publicly available.
\end{abstract}

\section{Introduction} \label{sec:intro}
 Early and accurate diagnosis of parkinsonian syndromes is critical to provide appropriate care to patients and for inclusion in therapeutic trials. Quantitative susceptibility mapping (QSM) provides a novel contrast mechanism that provides for indirect estimates brain iron levels in vivo~\cite{sjostrom2017quantitative}. The red nucleus and the dentate nucleus are small brain structures that are of particular interest for parkinsonian syndromes~\cite{williams2009progressive} and that can be studied with QSM. Automatic segmentation of these structures from QSM data could provide useful biomarkers of these neurodegenerative disorders.

Acquiring and labeling medical data is expensive, especially for segmentation tasks that require annotating every voxel of the target structures.
Efficient training with limited data remains a key challenge in medical imaging. This is particularly relevant for tasks where large annotated datasets are not available as it is the case in our application.

Xu et al.~\cite{xu2020frequency} and Rahamman et al. ~\cite{rahaman2019spectral} found that deep networks tend to fit from low to high frequency information during training. This phenomenon was referred to as the frequency principle (F-principle) of deep learning~\cite{xu2020frequency}.
The unbalanced learning of high- and low-frequency information during training requires a large amount of data to produce reliable results.
Tang et al. ~\cite{tang2022defects} analyzed from the frequency domain and proved that cascaded convolutional decoder networks are more likely to weaken high-frequency components.
From the above references, we can conclude that different frequencies play different roles in the learning process of deep networks.
Therefore, to effectively learn information, the learning process must balance between high- and low-frequency components.  Furthermore,  it has been shown that CNN decoders (which are a part of most deep learning segmentation methods) weaken the impact of high-frequency information during training ~\cite{tang2022defects}.  One may thus hypothesize that disentangling image information into high- and low-frequency components may lead to improved segmentation results.

Various recent works have proposed to exploit disentanglement in deep learning.  Azad et al. ~\cite{azad2021deep} proposed a frequency re-calibration U-Net for medical image segmentation,  by introducing the Laplacian pyramid in the U-shaped structure.  It allowed to  generalize better in low training data.
Liu et al. ~\cite{liu2020defending} used frequency refinement to improve adversarial defense for several biomedical image segmentation tasks.
McIntosh et al. ~\cite{mcintosh2021preservation} proposed a wavelet transform-based model and showed that extra high-frequency components can increase the performance.  
Charstias et al. ~\cite{chartsias2019disentangled} proposed to decompose  cardiac images into spatial anatomical factors and non-spatial modality factors using  a variational autoencoder.
Liu et al. ~\cite{liu2022disentangled} proposed a method for optical coherence tomography angiography (OCTA) segmentation based on disentangling images into the anatomy component and the local contrast component from paired OCTA scans.
The disentangling module is implemented by a conditional variational autoencoder (CVAE).
Furthermore,  several works have used frequency decomposition for domain adaptation and generalization ~\cite{liu2021feddg,yang2020fda,huang2021fsdr}.
However, the above approaches may be complex to train and implement.  Moreover, they may be specific to a given architecture. Finally, they did not specifically assess the impact in the low training size regime.

In this paper,  we propose a new medical image segmentation method using disentangled high- and low-frequency parts for segmentation of red and dentate nuclei from QSM data in Parkinson's disease.  The method has two strong advantages: it is conceptually simple and it can be used with any type of segmentation network.  In addition, we also evaluate the method on three tasks from the Medical Segmentation Decathlon (MSD) ~\cite{antonelli2022medical}. We specifically study the behavior of our method when training sets are very small.

\section{Methods} \label{sec:methods}
\begin{figure}[!hbtp]
\centering
\includegraphics[width=1\linewidth]{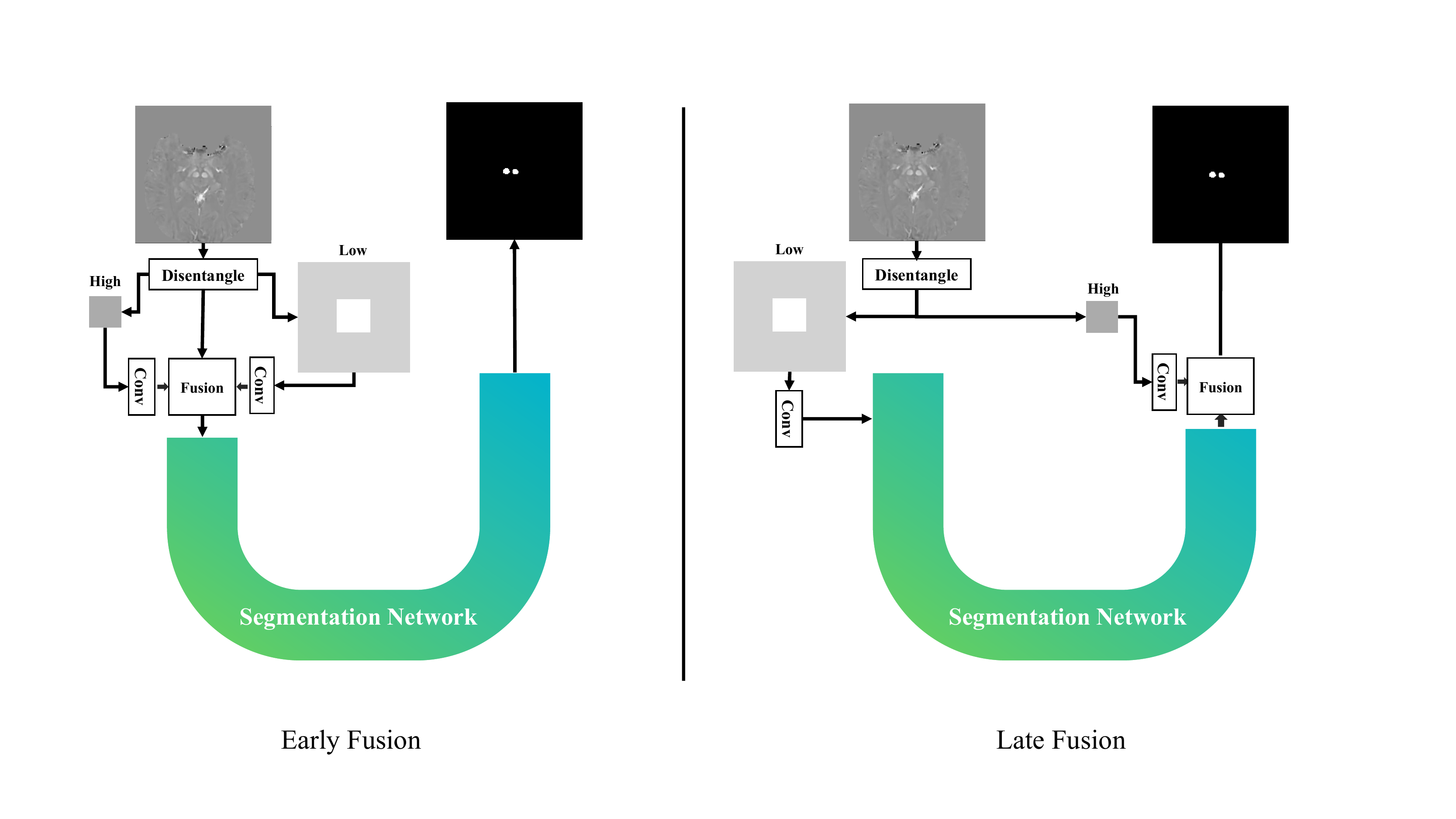}
\caption{Processing flow of the proposed method. We introduce two ways of fusion: early fusion and late fusion.}
\label{fig:overall_flow}
\end{figure}

The proposed method consists of two steps: i) frequency domain disentangling and feature learning; ii) frequency domain fusion. Two types of fusion are considered. In early fusion, the fusion is done before feeding the result to a segmentation network. In late fusion, only the high frequency information is fed to the segmentation network and the result is fused with low frequency learned features. Our method is general and can be applied to any segmentation network. The overall workflow of the approach can be seen in Figure~\ref{fig:overall_flow}.
The code for our method is available at \url{https://anonymous.4open.science/r/frequency_disentangled_learning-1C81}. 

\subsection{Frequency domain disentangling and feature learning}\label{sec:methods:disentangle}

Given samples $i \in I$, where $I \subset \mathbb{R}^{N_x \times N_y \times N_z}$ is a set of images,
the disentangling operation is achieved by first transferring to Fourier space, and separating the high- and low-frequency components as follows:
\begin{equation}
\begin{aligned}
& \mathcal{H}^\theta({i}) = \mathcal{F}({i})\left[\frac{N_x\times (1-\theta)}{2}: \frac{N_x\times(1+\theta)}{2},\frac{N_y\times(1-\theta)}{2}: \frac{N_y\times(1+\theta)}{2}, :\right]\\
&\mathcal{L}^\theta({i}) = \mathcal{F}({i}) - \mathcal{H}({i})
\end{aligned}
\label{eq:fourier_disentangle}
\end{equation}
where $\mathcal{F}(i)$ represents the Fourier transform of $i$, $\mathcal{L}^\theta(i)$ is the extraction of the low-frequency part of $i$, $\mathcal{H}^\theta(i)$ is the high-frequency part and $\theta \in (0,1)$ is a parameter that controls the high/low frequency separation.
We then apply the inverse Fourier transform ($\mathcal{F}^{-1}$) to obtain high- and low-frequency parts in image space:
\begin{equation}
\begin{aligned}
&L^{\theta}(i)=\mathcal{F}^{-1}(\mathcal{L}^{\theta}(i))\\
&H^{\theta}(i)=\mathcal{F}^{-1}(\mathcal{H}^{\theta}(i))\\
\end{aligned}
\end{equation}
$L^{\theta}(i)$ and $H^{\theta}(i)$ are then each fed to a convolutional layer and the outputs are respectively denoted as $O^{\theta}_{L}$ and $O^{\theta}_{H}$.

The F-principle indicates that models tend to first fit low-frequency information. Thus, in particular with low sample size, there is a risk that high-frequency information is not adequately learnt. High-frequency information represents structural details that are essential in medical tasks.
Our approach addresses this issue by utilizing a simple disentanglement operation that forces the model to balance the learning process for both high and low-frequency information.

\subsection{Feature fusion after disentangled learning} \label{sec:methods:feature_fusion}
The fusion operation can be done at two different stages: early fusion and late fusion. In the case of early fusion, low- and high-frequency outputs $O^{\theta}_{L}$ and $O^{\theta}_{H}$ are fused before being fed to a segmentation network. In the case of late fusion, only the high frequency is fed to a segmentation network, resulting in a result denoted as $S^{\theta}_{H}$ which is fused with $O^{\theta}_{L}$.

\begin{figure}[htbp]
\centering
\includegraphics[width=0.8\linewidth]{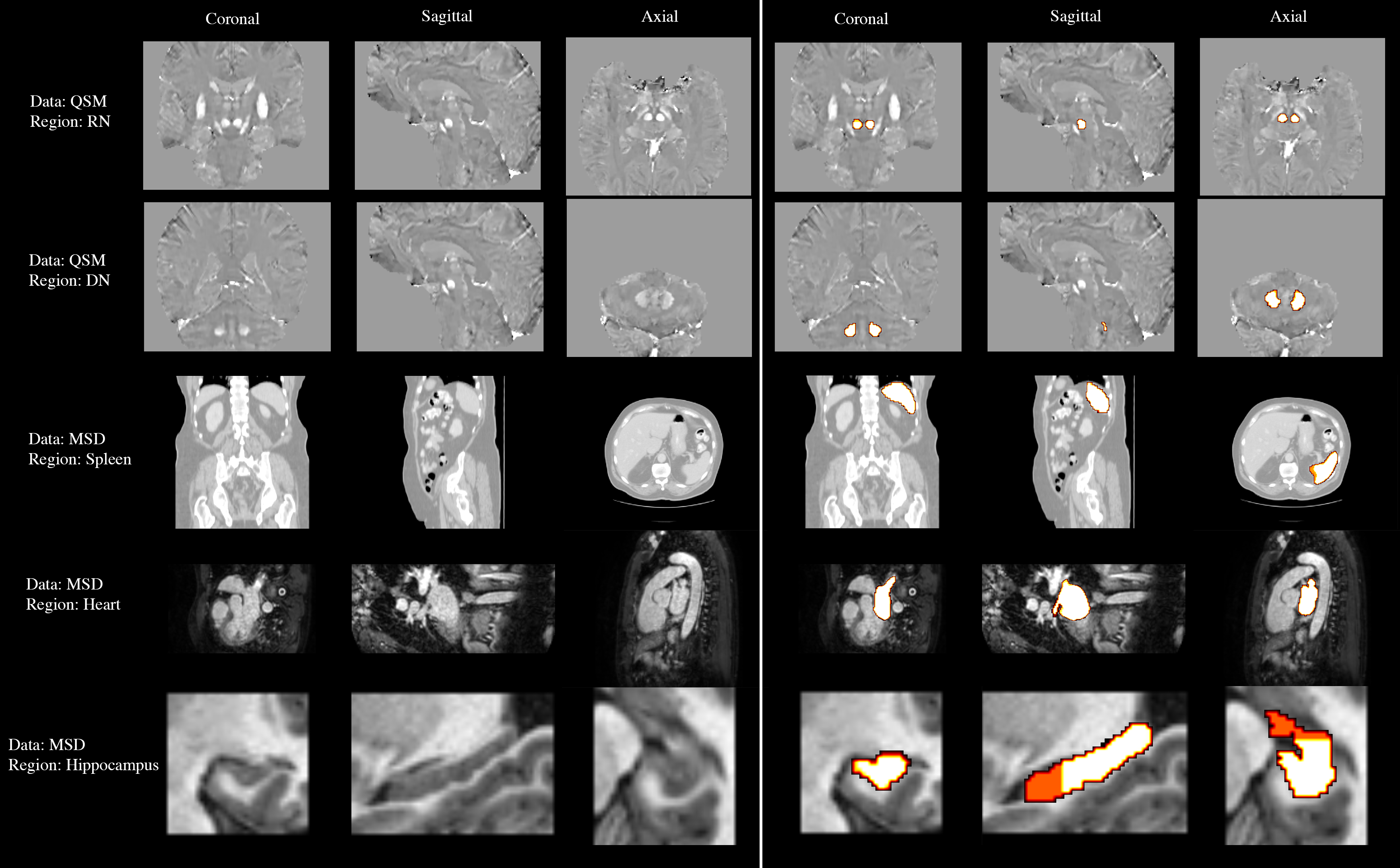}
\caption{These are examples of all the datasets we used for model evaluation. We overlap the region of interest (right part) and visualize them in three planes. }
\label{fig:dataset}
\end{figure}

\section{Experiments and results} \label{sec:experiments}

\subsection{Implementation details}
Our code is developed based on the PyTorch framework ~\cite{paszke2019pytorch}.
We used the open-source Python library TorchIO ~\cite{perez2021torchio} for reshaping images to the same size (for a given task) and for min-max normalization.
We used Adam ~\cite{kingma2014adam} as optimizer using an learning rate of 1e-3.
We did not apply any hyperparameter selection techniques or data augmentation.
In our experiments, we used a 3D-UNet ~\cite{cciccek20163d} as segmentation network and Dice as loss function ~\cite{milletari2016v} but the approach is general and could be applied to other models.

\subsection{Datasets}
We conducted experiments on five segmentation tasks: two tasks from a private dataset and three tasks from public datasets from the Medical Segmentation Decathlon (MSD) ~\cite{antonelli2022medical}. Each dataset was split into training validation and test sets. The splits were done at the participant level to avoid any data leakage ~\cite{thibeau2022clinicadl}.
Specifically, the models were trained with different percentage of the training set to explore the ability under small datasets, while the validation and test sets were left unchanged. The datasets and the splits are summarized in Table~\ref{tab:data_distribution}.
We used as performance metrics the Dice coefficient and the 95\% Hausdorff distance. For each metric, we report its mean value over the independent test set as well as the corresponding 95\% confidence interval (CI) computed using bootstrap.
We evaluated the performance of our approach for different sizes of the training set, while the validation and test sets were left unchanged.

Figure~\ref{fig:dataset} shows the data and the region of interest for each segmentation task.

\begin{table}[hbtp!]
\begin{center}
\caption{Characteristics of the 3D medical imaging datasets.}
\resizebox{1.0\linewidth}{!}{
\begin{tabular}{|l|l|l|l|l|l|}
\hline
Data type & Dataset                & Task                                 & Train+val & Test & Image Size \\
\hline
3D MRI    & {[}Private] QSM        & Red Nucleus  (RN)       & 51+13     & 16   & 160,160,128         \\
3D MRI    & {[}Private] QSM        & Dentate Nucleus  (DN)   & 42+11     & 14   & 160,160,128          \\
3D CT     & {[}Public] Spleen      & Full organ          & 25+7      & 9    & 160,160,128         \\
3D MRI    & {[}Public] Hippocampus & Anterior, Posterior & 166+42    & 52   & 64, 64, 48          \\
3D MRI    & {[}Public] Heart       & Full organ          & 12+4      & 4    & 160,160,128        \\
\hline
\end{tabular}
}
\label{tab:data_distribution}%
\end{center}
\end{table}

The private dataset comprised 18 healthy subjects, 46 patients with early Parkinson’s disease (i.e.  disease duration below 4 years), and 16 patients with prodromal parkinsonism (idiopathic rapid eye movement sleep behavior disorder-iRBD), recruited between May 2015 and January 2019 as part of the ICEBERG cohort.
The tasks were to segment the red nucleus (RN) and the dentate nucleus (DN) from QSM data. QSM were generated from multi-echo 3D GRE (12 echo times ranging from 4 ms to 37 ms) with a full brain coverage at an isotropic voxel resolution of 1~mm$^3$.
For RN segmentation, the training, validation and test sets respectively contain 51, 13 and 16 participants.
In some participants, the boundaries of the DN were heavily affected by artifacts, making them impossible to distinguish. Such participants were excluded from the DN segmentation task which included 67 participants (training, validation and test sets comprised 42, 11, and 14 participants, respectively).

The three tasks from MSD were spleen, heart and hippocampus segmentation.
The heart task is to segment the left atrium in a  mono-modal MRI scan.
The hippocampus task is to segment the anterior and posterior parts of the hippocampus based on MRI.
The spleen segmentation task is to segment  the spleen in the portal phase in a CT scan.

We studied the performance when varying the size of the training set, ranging from very small size (4 samples) to full training set (166 participants for the hippocampus task).

\begin{table}[hbt]
\centering
\caption{Performance on private dataset (average value [95\% bootstrap confidence interval]).  FD indicates whether frequency disentangled learning was performed; $n$ represents the size of the subsample used for training.}
\label{tab:performance_QSM}
\resizebox{0.6\linewidth}{!}{
\begin{tabular}{|c|l|l|l|l|}
\hline
Task-Region              & n (\%)                      & FD    & Dice (\%)           & 95 Hausdorff         \\
\hline
\multirow{15}{*}{QSM-RN} & \multirow{3}{*}{4 (7.5\%)}  & None  & 45.80 [40.21,51.62]  & 70.04 [67.18,72.84]  \\
                         &                             & Early & 80.60 [77.84,83.14]  & 8.88 [1.05,20.52]    \\
                         &                             & Late  & 82.95 [78.83,85.92] & 4.8 [1.03,12.28]     \\
\cline{2-5}
                         & \multirow{3}{*}{8 (15\%)}   & None  & 84.17 [81.85,86.27] & 5.66 [1.03,14.14]    \\
                         &                             & Early & 88.45 [87.31,89.37] & 4.57 [1.0,11.72]     \\
                         &                             & Late  & 83.73 [81.02,86.13] & 12.94 [2.61,25.92]   \\
\cline{2-5}
                         & \multirow{3}{*}{16 (30\%)}  & None  & 77.36 [74.79,79.7]  & 19.97 [8.03,32.88]   \\
                         &                             & Early & 87.28 [85.18,89.17] & 6.66 [1.0,15.06]     \\
                         &                             & Late  & 88.84 [87.64,90.12] & 1.0 [1.0,1.0]        \\
\cline{2-5}
                         & \multirow{3}{*}{26 (50\%)}  & None  & 87.57 [86.15,88.96] & 1.0 [1.0,1.0]        \\
                         &                             & Early & 89.78 [88.66,90.84] & 1.0 [1.0,1.0]        \\
                         &                             & Late  & 89.76 [88.75,90.69] & 1.0 [1.0,1.0]        \\
\cline{2-5}
                         & \multirow{3}{*}{51 (100\%)} & None  & 88.21 [86.74,89.72] & 1.0 [1.0,1.0]        \\
                         &                             & Early & 89.90 [89.04,90.74]  & 1.0 [1.0,1.0]        \\
                         &                             & Late  & 90.58 [88.92,91.84] & 4.87 [1.0,12.6]      \\
\hline
\multirow{15}{*}{QSM-DN} & \multirow{3}{*}{4 (7.5\%)}  & None  & 51.09 [44.37,56.82] & 59.4 [46.83,71.55]   \\
                         &                             & Early & 79.59 [76.41,82.72] & 2.16 [1.47,3.05]     \\
                         &                             & Late  & 78.82 [75.74,81.73] & 8.3 [4.11,13.39]     \\
\cline{2-5}
                         & \multirow{3}{*}{7 (15\%)}   & None  & 69.23 [63.03,74.68] & 37.65 [21.25,55.24]  \\
                         &                             & Early & 81.44 [78.49,84.2]  & 7.77 [2.39,14.53]    \\
                         &                             & Late  & 81.89 [78.85,84.64] & 2.93 [1.29,5.45]     \\
\cline{2-5}
                         & \multirow{3}{*}{13 (30\%)}  & None  & 75.63 [70.83,79.99] & 11.23 [2.12,25.38]   \\
                         &                             & Early & 82.65 [79.08,85.75] & 5.22 [1.15,12.51]    \\
                         &                             & Late  & 84.00 [81.21,86.44]  & 1.44 [1.06,2.01]     \\
\cline{2-5}
                         & \multirow{3}{*}{21 (50\%)}  & None  & 77.72 [73.71,81.38] & 11.94 [1.93,23.69]   \\
                         &                             & Early & 83.56 [79.76,86.74] & 3.72 [1.5,6.44]      \\
                         &                             & Late  & 82.11 [78.0,85.45]  & 2.46 [1.26,4.06]     \\
\cline{2-5}
                         & \multirow{3}{*}{42 (100\%)} & None  & 64.56 [58.95,70.05] & 49.7 [32.24,65.95]   \\
                         &                             & Early & 83.61 [79.38,87.24] & 3.41 [1.09,7.71]     \\
                         &                             & Late  & 85.73 [81.48,88.82] & 2.98 [1.0,6.71]      \\
\hline
\end{tabular}
}
\end{table}

\begin{table}[htb]
\centering
\caption{Performance on public datasets (average value [95\% bootstrap confidence interval]).  FD indicates whether frequency disentangled learning was performed; $n$ represents the size of the subsample used for training.}
\label{tab:performance_decathlon}
\resizebox{0.6\linewidth}{!}{

\begin{tabular}{|c|l|l|l|l|}
\hline
Task-Region                   & n (\%)                       & FD    & Dice (\%)           & 95 Hausdorff           \\
\hline
\multirow{9}{*}{Spleen}       & \multirow{3}{*}{4 (15\%)}    & None  & 64.29 [52.44,76.00]  & 129.9 [65.43,194.98]   \\
                              &                              & Early & 69.04 [61.26,76.94] & 90.76 [37.05,151.65]   \\
                              &                              & Late  & 68.15 [58.02,78.58] & 116.52 [53.75,179.28]  \\
\cline{2-5}
                              & \multirow{3}{*}{8 (30\%)}    & None  & 72.99 [62.57,83.06] & 119.67 [46.57,197.53]  \\
                              &                              & Early & 77.81 [71.11,84.69] & 51.71 [22.12,85.3]     \\
                              &                              & Late  & 80.38 [73.57,86.93] & 41.99 [11.92,85.23]    \\
\cline{2-5}
                              & \multirow{3}{*}{25 (100\%)}  & None  & 85.50 [79.82,90.33]  & 29.83 [4.93,66.79]     \\
                              &                              & Early & 87.28 [82.63,91.24] & 29.58 [6.72,66.66]     \\
                              &                              & Late  & 86.78 [78.08,92.58] & 43.96 [5.63,96.68]     \\
\hline
\multirow{6}{*}{Heart}        & \multirow{3}{*}{4 (30\%)}    & None  & 79.74 [73.13,85.43] & 22.31 [8.22,48.52]     \\
                              &                              & Early & 82.57 [77.35,87.95] & 20.16 [5.48,45.18]     \\
                              &                              & Late  & 82.92 [76.57,87.59] & 8.72 [5.91,13.24]      \\
\cline{2-5}
                              & \multirow{3}{*}{12 (100\%)}  & None  & 83.48 [80.18,87.29] & 18.46 [5.01,32.66]     \\
                              &                              & Early & 87.73 [84.13,91.40]  & 7.18 [2.68,11.68]      \\
                              &                              & Late  & 88.43 [86.39,90.47] & 6.34 [2.85,12.09]      \\
\hline
\multirow{18}{*}{Hippocampus} & \multirow{3}{*}{4 (2.5\%)}   & None  & 74.69 [73.15,76.14] & 3.4 [2.78,4.11]        \\
                              &                              & Early & 75.98 [74.6,77.25]  & 3.45 [2.79,4.21]       \\
                              &                              & Late  & 76.48 [75.08,77.77] & 2.75 [2.36,3.27]       \\
\cline{2-5}
                              & \multirow{3}{*}{8 (5\%)}     & None  & 78.67 [77.32,79.91] & 2.58 [2.14,3.12]       \\
                              &                              & Early & 79.78 [78.50,80.91]  & 3.38 [2.49,4.38]       \\
                              &                              & Late  & 79.90 [78.51,81.13]  & 3.09 [2.36,3.96]       \\
\cline{2-5}
                              & \multirow{3}{*}{17 (10\%)}   & None  & 80.04 [78.44,81.44] & 2.15 [1.86,2.5]        \\
                              &                              & Early & 81.14 [79.46,82.6]  & 1.91 [1.65,2.22]       \\
                              &                              & Late  & 81.06 [79.63,82.31] & 1.97 [1.71,2.27]       \\
\cline{2-5}
                              & \multirow{3}{*}{50 (30\%)}   & None  & 82.41 [81.23,83.47] & 1.63 [1.47,1.83]       \\
                              &                              & Early & 83.31 [82.31,84.2]  & 1.47 [1.34,1.62]       \\
                              &                              & Late  & 82.95 [81.96,83.87] & 1.71 [1.44,2.08]       \\
\cline{2-5}
                              & \multirow{3}{*}{83 (50\%)}   & None  & 84.17 [83.28,85.01] & 1.41 [1.3,1.55]        \\
                              &                              & Early & 84.59 [83.73,85.37] & 1.35 [1.25,1.46]       \\
                              &                              & Late  & 84.30 [83.24,85.22]  & 1.39 [1.27,1.52]       \\
\cline{2-5}
                              & \multirow{3}{*}{166 (100\%)} & None  & 85.82 [84.72,86.77] & 1.38 [1.23,1.56]       \\
                              &                              & Early & 85.89 [84.97,86.75] & 1.33 [1.22,1.45]       \\
                              &                              & Late  & 86.26 [85.49,87.0]  & 1.25 [1.16,1.36]       \\
\hline
\end{tabular}
}
\end{table}

\subsection{Results}
Results for DN and RN from QSM data are presented in Table~\ref{tab:performance_QSM}. For the DN, our approach resulted in considerable improvements in performance, even when using the full dataset for training (e.g. Dice of 83-85 vs 64 and non-overlapping CIs). For the DN, there was no major improvement when using the full dataset. On the other hand, our methods resulted in major improvements when the training set was small (7.5\%, 15\% and 30\% of training data).

Results of spleen, heart and hippocampus segmentation are shown in Table~\ref{tab:performance_decathlon}. One can notice substantial (up to 8 points of Dice) and systematic improvements for the spleen and heart datasets even though the CIs are overlapping (due to small test sets).
For the hippocampus, frequency disentanglement did not provide any substantial improvement.

\section{Discussion} \label{sec:discussion}

We presented a novel segmentation approach based on disentangling of frequency components. Applied to the segmentation of the red and dentate nuclei from QSM, it provided considerable improvements in performance.  

Our method is grounded on the frequency principle of deep learning.  Conventional training process can result in asynchronous learning in the frequency domain. Our solution is to disentangle the data without the need to modify the model architecture or employ other training techniques.
This method has the benefit of being simple to implement and applicable with any segmentation network.  In our experiments,  we used a U-Net as backbone as it is arguably the most standard segmentation approach.  Further work could study the benefit of our approach with more complex networks.  

We propose two fusion strategies: early and late.   In our experiments,  we did not observe any major or systematic difference in performance between the two. However,  the early fusion strategy offers more flexibility and is not restricted to an encoder-decoder architecture.  

Our approach provided considerable improvements in performance for RN and DN segmentation, in particular for small sample size.  On the MSD data,  results were mixed: the improvement was substantial for the heart and spleen but not for the hippocampus.  A strength of our study is that we systematically computed bootstrap confidence intervals which, unlike providing standard-deviation over cross-validation folds~\cite{bengio2004no},  is a statistically rigorous way to study the precision of the performance estimates. However,  a limitation is that our testing sets were often small which led to wide confidence intervals (the width roughly varies as a multiple of $1/\sqrt{n}$)~\cite{eljurdi2023}.

    Another limitation of our approach is that one needs to choose the parameter $\theta$ which controls the separation between high and low frequencies. We did not experiment with varying values of $\theta$. It is possible that other values would have been more adapted for some tasks. Future work could aim to integrate the parameter into the loss function as  in reference~\cite{cai2021frequency}.

\section{Compliance with ethical standards}
The institutional ethical standard committee approved the study (CPP Paris VI/RCB: 2014-A00725-42).
All participants gave written informed consent.

\section{Acknowledgments}\label{sec:acknowledgments}
The research leading to these results has received funding from the French government under management of Agence Nationale de la Recherche (ANR) as part of the "Investissements d'avenir" program, references ANR-19-P3IA-0001 (PRAIRIE 3IA Institute) and ANR-10-IAIHU-06.
The ICEBERG study is supported by the European Research Council (ERC) (No. 678304),  the Horizon 2020 (No. 826421-TVB-Cloud),  ANR (ANR-11-INBS-0006,  ANR-19-JPW2-000),  France Parkinson (PRECISE-PD project), the Fondation d’Entreprise EDF, Biogen Inc.,  the Fondation Thérèse and René Planiol, the Fondation Saint Michel,  Energipole (M. Mallart), M. Villain and the Société Française de Médecine Esthétique (M. Legrand).
Guanghui Fu is supported by the Chinese Government Scholarship provided by China Scholarship Council (CSC). Lydia Chougar is supported by a Poste d'accueil Inria/AP-HP.

\bibliographystyle{splncs04}
\bibliography{refs}
\end{document}